\documentclass[aps,prl,reprint,superscriptaddress,nofootinbib,floatfix]{revtex4-2}

\usepackage{amsmath,amssymb,mathtools,bm}
\usepackage{graphicx}
\usepackage{xcolor}
\usepackage[colorlinks=true,citecolor=blue,linkcolor=blue,urlcolor=blue]{hyperref}
\usepackage[normalem]{ulem}
\newcommand{\Tr}{\operatorname{Tr}}
\newcommand{\sech}{\operatorname{sech}}

\newcommand{\Ds}{D_s}
\newcommand{\Rone}{R_1}
\newcommand{\FO}{F_\mathcal{O}}
\newcommand{\ii}{\mathrm{i}}

\begin{document}

\title{Quantum Fisher Information Deficit as a Geometric
Signature of \\
Strong-to-Weak Spontaneous Symmetry Breaking}

\author{Zhang-Chi Yin}
\affiliation{Hefei National Research Center for Physical Sciences at the Microscale and School of Physical Sciences,
University of Science and Technology of China,
Hefei 230026, China}
\author{Wei Zheng}
\email{zw8796@ustc.edu.cn}
\affiliation{Hefei National Research Center for Physical Sciences at the Microscale and School of Physical Sciences,
University of Science and Technology of China,
Hefei 230026, China}
\affiliation{CAS Center for Excellence in Quantum Information and Quantum Physics,
University of Science and Technology of China, Hefei 230026, China}
\affiliation{Hefei National Laboratory,
University of Science and Technology of China, Hefei 230088, China}

\begin{abstract}
Recently, strong-to-weak spontaneous symmetry breaking (SW-SSB) has been established as a framework for describing quantum phases in open systems. However SW-SSB cannot be characterized by conventional charged two-point correlators. In this work, we show that, for strongly $\mathbb{Z}_2$-symmetric states with local charged involutions, SW-SSB manifests as a persistent deficit from the maximal quantum Fisher information (QFI) under symmetry-neutral pair rotations. A model-independent inequality hierarchy establishes the equivalence of this deficit with fidelity and Rényi-1 correlators as asymptotic long-range-order diagnostics. It thereby connects finite-angle state overlaps to a constraint on the infinitesimal Bures geometry along the same pair-generated orbit. For symmetry-projected Gibbs states, we further derive a lower bound on the deficit governed by the mean energy injected by the pair operation. Crucially, locality keeps this energy cost uniformly bounded in system size and endpoint separation, maintaining a finite gap below the maximal QFI at any fixed finite temperature. When ordinary charged long range order is absent, this persistent geometric constraint provides a quantitative information-geometric characterization of thermal SW-SSB.
\end{abstract}
\maketitle
\paragraph*{Introduction.---}Spontaneous symmetry breaking in open quantum systems unveils rich ordering phenomena beyond pure state paradigms and represents an active frontier of nonequilibrium many-body physics~\cite{diehl2008,sieberer2025,lee2023,gu2025,wangreview2026}. Distinct from isolated pure states, mixed states described by a density matrix $\rho$ admit two inequivalent notions of symmetry under a group representation $\hat{U}_g$. Specifically, $\rho$ exhibits strong symmetry when its support lies entirely within a definite charge sector obeying $\hat{U}_g\rho = e^{\ii\theta_g}\rho$, whereas weak symmetry requires only invariance under the adjoint action $\hat{U}_g\rho \hat{U}_g^\dagger = \rho$. This distinction gives rise to strong-to-weak spontaneous symmetry breaking (SW-SSB), a mixed-state phenomenon in which strong symmetry is spontaneously broken while weak symmetry remains intact~\cite{sala2024,lessa2025}. Formulated theoretically via canonical purifications and bilayer frameworks~\cite{sala2024,lubilayer2024,liu2025}, SW-SSB now underpins a broad class of intrinsic mixed-state phenomena, ranging from noisy circuit dynamics~\cite{kuno2024,singh2026} and decohered quantum criticality~\cite{lee2023,zou2023,guocritical2025} to mixed-state topological phases~\cite{leespt2025,guoyang2025,zhang1form2024,ellisoncheng2025}.

In an SW-SSB phase, long-range charged correlations can cancel when averaged over a mixed state, causing ordinary two-point correlators to vanish. The fidelity correlator $F_{\mathcal O}$~\cite{lessa2025} and the R\'enyi-1 (Wightman) correlator $R_1$~\cite{liu2025,weinstein2025} provide robust diagnostics of this hidden order but are difficult to measure in generic mixed states~\cite{feng2025}. In contrast, the R\'enyi-2 correlator $R_2$~\cite{sala2024,kuno2024} can be accessed through randomized measurements~\cite{sun2025}, but its long range order (LRO) is not generally stable under local quantum channels and may differ from fidelity phase boundaries~\cite{lessa2025}. These limitations motivate a complementary approach to probing SW-SSB through quantum response~\cite{zhang2024}.

Quantum Fisher information (QFI) characterizes the distinguishability of a quantum state under an infinitesimal transformation~\cite{wootters1981,braunstein1994,petzghinea2011} and has been widely employed in quantum metrology ~\cite{giovannetti2006,pezze2009,toth2014,pezze2018} as well as in the study of many-body phase~\cite{zanardi2007,venutizanardi2007,zanardi2008,hauke2016,pezze2017}. Recent work has further exploited QFI to triditional spontaneous symmetry breaking~\cite{yamashika2025,chengreen2026}. A natural question therefore arises as to whether QFI can effectively detect SW-SSB and corresponding phase transitions in open quantum systems.

\begin{figure}[t]
\includegraphics[width=0.48\textwidth]{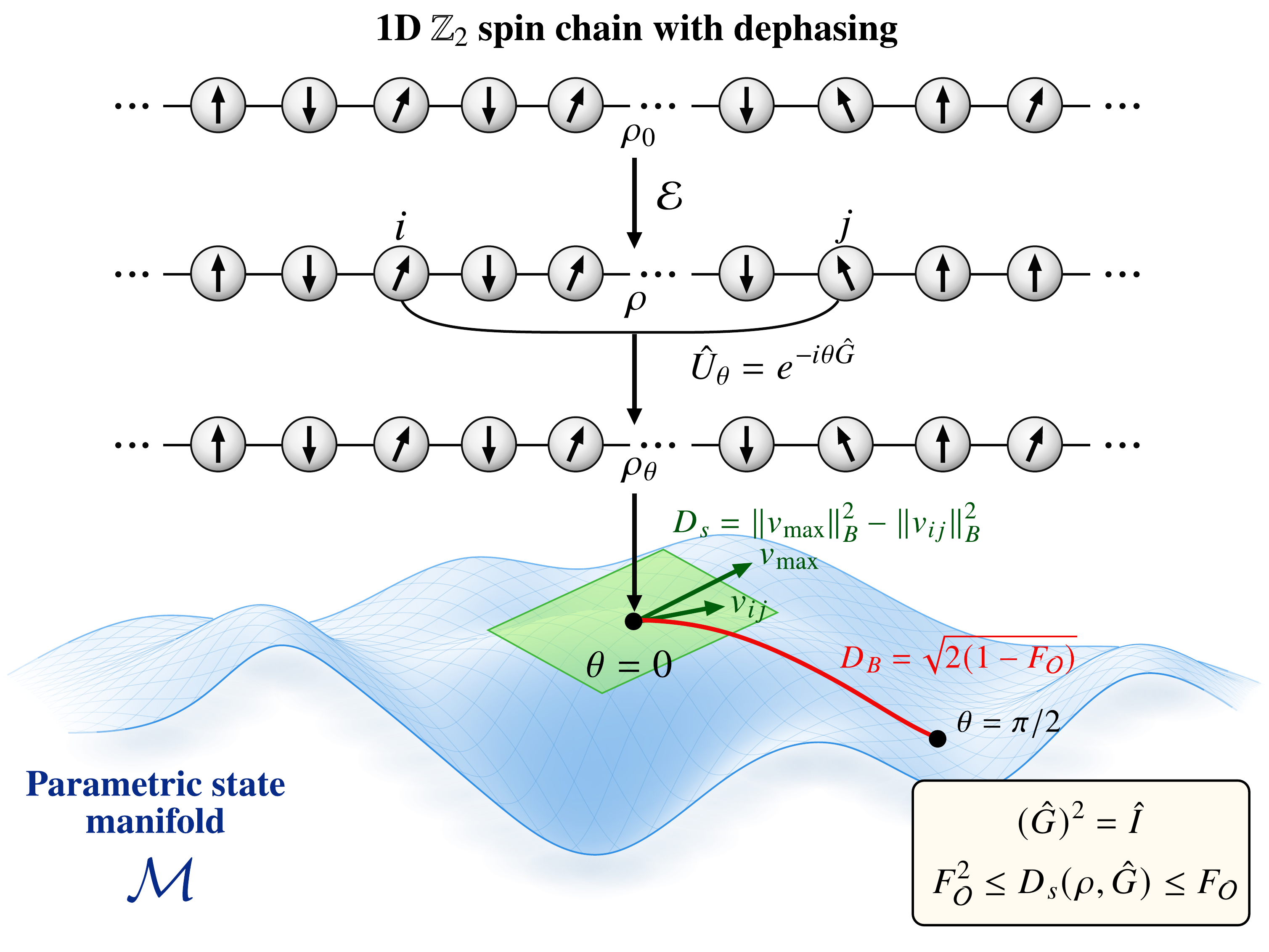}
  \caption{\label{fig:concept}Geometry of a neutral pair rotation in a dephased $\mathbb Z_2$ spin chain. The dephasing channel $\mathcal E$ maps $\rho_0$ to $\rho$, and the marked sites $i,j$ specify the generator $\hat{G}$. At $\theta=0$, the green arrows show the orbit tangent $v_{ij}$ and a reference tangent $v_{\max}$ with maximal Bures norm. The red curve represents the Bures distance $D_B$ between $\rho$ and $\rho_{\pi/2}$. We establish a quantitative relation between the QFI deficit $D_s$ and the fidelity correlator $F_{\mathcal O}$.}
\end{figure}

Conventionally, symmetry breaking is diagnosed through an enhanced or diverging response. In this Letter, we address whether this response paradigm extends to the strong-to-weak setting by subjecting the open-system state to a
$\theta$-parametrized, symmetry-preserving pair rotation [Fig.~\ref{fig:concept}]. For strongly $\mathbb{Z}_2$-symmetric states with local charged involutions, we show that SW-SSB manifests not as an enhancement of QFI with respect to the pair rotation, but as a persistent geometric deficit relative to its algebraic maximum in the long-distance limit. We then establish a model-independent inequality hierarchy that rigorously connects this QFI deficit, $D_s$, to the fidelity and Rényi-1 correlators, $F_\mathcal{O}^2 \le D_s \le R_1 \le F_\mathcal{O}$. This two-sided bound implies that the three quantities yield equivalent LRO criteria for SW-SSB, while remaining generically distinct at finite separation. Our findings thus provide a quantum-geometric characterization of SW-SSB LRO.

Furthermore, we extend this framework to symmetry-projected Gibbs states at finite temperature. Hamiltonian locality ensures that the energy cost of the pair rotation remains bounded independently of system size and endpoint separation, which yields a strictly positive lower bound on the QFI deficit governed by the mean injected energy, $D_s \ge e^{-\beta\Delta E_G}$. Consequently, at any fixed nonzero temperature, the QFI remains uniformly bounded away from its algebraic maximum, establishing a finite geometric gap that persists at arbitrary distances. In the absence of conventional charged LRO, this provides an information-geometric characterization of thermal SW-SSB, as conjectured by Lessa et al~\cite{lessa2025}.

\paragraph*{Quantum-geometric hierarchy.---}
We consider a $\mathbb{Z}_2$ spin chain initialized in a strongly symmetric state $\rho_0$. The probe state $\rho=\mathcal E(\rho_0)$ is prepared through a quantum channel or open-system evolution that preserves strong symmetry [Fig.~\ref{fig:concept}]. To probe SW-SSB in $\rho$, we introduce a local charged Hermitian
involution $\hat{O}$ satisfying $\hat{O}^2=\hat{I}$ and
$\hat{U}_g\hat{O}\hat{U}_g^\dagger=-\hat{O}$.
Combining two such operators on disjoint supports at $i$ and $j$ defines
the neutral Hermitian involution $\hat{G}\equiv\hat{O}_i\hat{O}_j^\dagger$.
This pair operator in turn generates a unitary orbit parametrized by the
rotation angle $\theta$ within the charge sector of $\rho$,
\begin{equation}
    \rho_\theta = e^{-\ii\theta \hat{G}}\rho \, e^{\ii\theta \hat{G}}.
    \label{eq:rotation}
\end{equation}
We denote the symmetric logarithmic derivative (SLD) QFI with respect to $\theta$ by $Q\equiv Q(i,j)=Q[\rho;\hat{G}]$. It determines the Bures line element $ds_B^2=(Q/4)d\theta^2$ at $\theta=0$. Since $\hat{G}^2=\hat{I}$ gives $Q(i,j)\le4$, we define the normalized QFI deficit as
\begin{align}
    D_s(i,j) \equiv 1 - \frac{Q(i,j)}{4}.
    \label{eq:definition}
\end{align}

Consider an ensemble $\mathcal A=\{w_k,|\psi_k\rangle\}$ of normalized pure states with probabilities $w_k$, realizing $\rho=\sum_k w_k|\psi_k\rangle\langle\psi_k|$. The pair expectation $g_k=\langle\psi_k|\hat{G}|\psi_k\rangle$ gives a generator variance $1-g_k^2$ in each branch. The QFI convex-roof identity, conjectured by T\'oth and Petz~\cite{tothpetz2013} and proved in general by Yu~\cite{yu2013}, states that $Q[\rho;\hat{G}]=4\min_{\mathcal A}\sum_k w_k(1-g_k^2)$, where the minimum runs over all decompositions of $\rho$. Equation~\eqref{eq:definition} therefore gives
\begin{equation}
  D_s = \max_{\mathcal A}\sum_k w_k g_k^2.
  \label{eq:ensembledeficit}
\end{equation}
In each pure-state branch, a nonzero expectation $g_k$ suppresses the sensitivity of $|\psi_k\rangle$ to the pair rotation by reducing the generator variance $1-g_k^2$. When $|g_k|=1$, $|\psi_k\rangle$ becomes an eigenstate of $\hat{G}$ with vanishing variance, as the rotation changes only its overall phase. Across the ensemble, positive and negative branch expectations can cancel in the linear correlator $\Tr(\rho\hat{G})=\sum_k w_k g_k$, while the squared average $\sum_k w_k g_k^2$ remains positive. Since Eq.~\eqref{eq:ensembledeficit} maximizes this squared average over all ensemble decompositions of $\rho$, the existence of such a decomposition guarantees a positive QFI deficit.

We compare this QFI deficit with two established nonlinear correlators: the root-fidelity correlator $F_{\mathcal O}(i,j)=\Tr\sqrt{\sqrt{\rho}\hat{G}\rho\hat{G}\sqrt{\rho}}$ ~\cite{uhlmann1976,jozsa1994} and the R\'enyi-1 (Wightman) correlator $R_1(i,j)=\Tr(\sqrt{\rho}\hat{G}\sqrt{\rho}\hat{G})$~\cite{luozhang2004,liu2025}.Here, $\mathcal O$ denotes the chosen family of charged endpoint operators. These correlators quantify overlaps between $\rho$ and its finite-angle image $\rho_{\pi/2}=\hat{G}\rho\hat{G}$, whereas $Q$ measures the local geometry along the same unitary orbit.

Purification and ensemble parametrization~\cite{uhlmann1976,hjw1993} similarly express the fidelity correlator as $F_\mathcal{O}=\max_{\mathcal A}\sum_k w_k|g_k|$ over all pure-state decompositions of $\rho$. For an ensemble maximizing $F_\mathcal{O}$, the Cauchy--Schwarz inequality bounds $F_\mathcal{O}^2$ by the weighted average of $g_k^2$ over that ensemble. Since $D_s$ is the maximum of this average over all pure-state decompositions [Eq.~\eqref{eq:ensembledeficit}], we obtain $F_\mathcal{O}^2\le D_s$. Moreover, to relate $D_s$ and $R_1$, we expand $\rho=\sum_a\lambda_a|a\rangle\langle a|$ in its eigenbasis and set $G_{ab}=\langle a|\hat{G}|b\rangle$. In this spectral representation, $D_s = \sum_{a,b:\lambda_a+\lambda_b>0} \frac{2\lambda_a\lambda_b}{\lambda_a+\lambda_b}|G_{ab}|^2$ weights each squared matrix element by the harmonic mean of the two eigenvalues, whereas $R_1 = \sum_{a,b}\sqrt{\lambda_a\lambda_b}|G_{ab}|^2$ uses their geometric mean. Comparing these weights term by term gives $D_s\le R_1$, which expresses the standard relation between SLD QFI and Wigner--Yanase skew information in terms of the deficit and overlap~\cite{luo2003,luo2004,petzghinea2011}. Finally, trace-norm duality gives $R_1\le F_\mathcal{O}$~\cite{audenaert2007,wilde2018,supplement}.

These three inequalities establish our first main result, valid for any density matrix $\rho$ and Hermitian involution $\hat{G}$,
\begin{equation}
  F_\mathcal{O}^2 \le D_s(\rho,\hat{G}) \le R_1 \le F_\mathcal{O}.
  \label{eq:chain}
\end{equation}
To apply this hierarchy to spatial LRO, we fix the local endpoint operator $\hat{O}$ and consider a sequence of bulk endpoint pairs separated by $r=d(i,j)$.Let $X^{(L)}(r)$ denote any of the three diagnostics in a finite system of size $L$.The thermodynamic limit at fixed separation defines $X(r) \equiv \lim_{L\to\infty} X^{(L)}(r)$.Taking the endpoint separation to infinity then defines the LRO parameter$\mathcal{L}[X] \equiv \lim_{r\to\infty} X(r)$.Provided these successive limits exist, Eq.~\eqref{eq:chain} implies
\begin{equation}
  \mathcal{L}[F_{\mathcal O}]>0 \Longleftrightarrow \mathcal L[D_s]>0 \Longleftrightarrow \mathcal L[R_1]>0.
  \label{eq:lro}
\end{equation}
Equation~\eqref{eq:lro} establishes equivalent LRO criteria for $D_s$, $F_{\mathcal O}$, and $R_1$. Consequently, unlike the R\'enyi-2 correlator, $D_s$ shares the same critical points as $F_{\mathcal O}$ and $R_1$ and inherits their stability under strongly symmetric finite-depth local quantum channels~\cite{lessa2025,liu2025,weinstein2025}. For a symmetric mixed state whose ordinary charged correlator vanishes in the long-distance limit, a nonvanishing long-range QFI deficit $D_s$ directly diagnoses SW-SSB in the chosen operator channel.
\begin{figure*}[t]
  \centering
  \includegraphics[width=\textwidth]{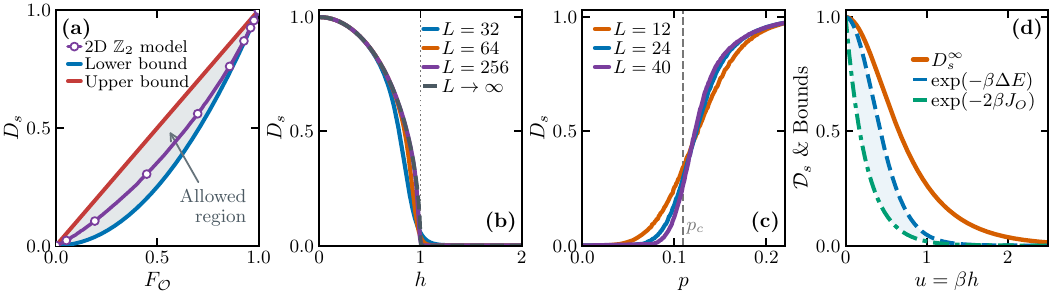}
  \caption{\label{fig:summary} Geometric bounds and model realizations.
  (a) Allowed physical region in the $(F_\mathcal{O}, D_s)$ plane bounded by $D_s = F_\mathcal{O}$ (solid red) and $D_s = F_\mathcal{O}^2$ (solid blue). Purple open circles trace a representative finite-size trajectory ($L=24$) for the two-dimensional model.
  (b) ZXZ deficit versus the preparation field $h$ after complete even-sublattice $\hat{X}$ dephasing, for $L=32,64,256$ and $r=L/2$. Solid curves show finite-size free-fermion results; the gray dashed curve is the thermodynamic analytic reference, and the vertical dotted line marks $h=1$.
  (c) Finite-size QFI deficit $D_s$ versus edge-dephasing probability $p$ for the two-dimensional $\mathbb Z_2$ model on open square lattices with $L=12,24,40$ and endpoint separation $r=L/2$. Distinct solid lines join sampled means. The gray dashed line marks the literature Nishimori point, $p_c=0.1092212(4)$~\cite{wan2026}, not a fit to these finite-size data.
  (d) Thermal QFI deficit $D_s = \sech(2u)$ (solid orange) for parity-projected noninteracting spins as a function of $u=\beta h$, compared against the injected-energy bound $e^{-4u\tanh u}$ (dashed blue) and the operator-norm bound $e^{-4u}$ (dash-dotted green). The shaded band highlights the quantitative tightening provided by the energy constraint.}
\end{figure*}

\paragraph*{Thermal energy bound.---}
To apply the above results to thermal SW-SSB, we consider the symmetry-projected Gibbs state $\rho_{\beta,P}=\hat{P}e^{-\beta \hat{H}}/Z_P$ at inverse temperature $0<\beta<\infty$~\cite{lessa2025,liu2025,weinstein2025}, where $\hat{P}$ is a nonzero orthogonal projector onto a definite charge sector satisfying $[\hat{P},\hat{H}]=0$ and $Z_P=\Tr(\hat{P}e^{-\beta \hat{H}})$. Acting within this sector ($[\hat{G},\hat{P}]=0$), the pair involution $\hat{G}$ injects a mean energy $\Delta E_G\equiv\Tr[\rho_{\beta,P}(\hat{G}\hat{H}\hat{G}-\hat{H})]$, which directly bounds the QFI deficit from below as our second main result:
\begin{equation}
  D_s(\rho_{\beta,P},\hat{G}) \ge e^{-\beta\Delta E_G}.
  \label{eq:thermalbound}
\end{equation}

To prove this bound, we introduce the Lehmann probability measure for the pair-induced spectral contributions. Specifically, this measure is built from pairs of energy eigenstates $|b\rangle$ and $|a\rangle$ connected by $\hat{G}$; each pair carries energy transfer $\omega=E_a-E_b$, Gibbs weight $\lambda_b=e^{-\beta E_b}/Z_P$.Multiplying the Gibbs weight by the squared matrix element and summing over all such pairs gives $d\mu_G(\omega)=\sum_{a,b}\lambda_b|G_{ab}|^2\delta\!\left[\omega-(E_a-E_b)\right]d\omega$, which is normalized by $\hat{G}^2=\hat{I}$ . Under this measure, the mean injected energy is given by the first moment, $\Delta E_G=\int\omega\,d\mu_G(\omega)$, while the Rényi-1 correlator is given by the exponential moment, $R_1=\int e^{-\beta\omega/2}d\mu_G(\omega)$. Jensen's inequality then yields $R_1\ge e^{-\beta\Delta E_G/2}$. Combining this estimate with the hierarchy in Eq.~\eqref{eq:chain}, which implies $D_s\ge R_1^2$, establishes Eq.~\eqref{eq:thermalbound}.

Beyond this bound, the same Lehmann measure also determines the thermal QFI through its spectral representation~\cite{hauke2016}
\begin{equation}
  Q = 4\int \tanh^2\!\left(\frac{\beta\omega}{2}\right) d\mu_G(\omega),
  \label{eq:FQ_spectral}
\end{equation}
Moreover, the deficit admits the spectral form $D_s=\int\sech^2(\beta\omega/2)\,d\mu_G(\omega)$. At any fixed finite temperature, the kernel $\sech^2(\beta\omega/2)$ is strictly positive on every bounded interval $|\omega|\le\Omega$ while decaying exponentially at large $|\omega|$. Consequently, $D_s$ can vanish in the thermodynamic limit only if the spectral weight within every finite frequency window vanishes, shifting entirely to unboundedly large $|\omega|$.Consistent with Eq.~\eqref{eq:thermalbound}, a vanishing deficit $D_s\to 0$ therefore requires a diverging mean injected energy $\Delta E_G\to\infty$.

For a local Hamiltonian $\hat{H}_L$, however, local endpoint commutators keep this energy cost finite. Assume $J_O\equiv\sup_{L,\hat{O}\in\mathcal O}\|[\hat{H}_L,\hat{O}]\|<\infty$, where the supremum runs over all system sizes and site copies of the endpoint operator. Because the endpoint operators have unit norm, the commutator product rule bounds $\|[\hat{H}_L,\hat{G}]\| \le \|[\hat{H}_L,\hat{O}_i]\|+\|[\hat{H}_L,\hat{O}_j]\| \le 2J_O$, which restricts the mean injected energy to $\Delta E_G=\Tr[\rho_{\beta,P}\,\hat{G}[\hat{H}_L,\hat{G}]] \le \|[\hat{H}_L,\hat{G}]\| \le 2J_O$. Substituting this estimate into Eq.~\eqref{eq:thermalbound} yields
\begin{equation}
  D_s(\rho_{\beta,P},\hat{G}) \ge e^{-2\beta J_O} > 0,
  \label{eq:localbound}
\end{equation}
which holds uniformly in system size $L$ and separation $r$ at any fixed finite temperature. From a quantum-geometric perspective, this persistent Bures deficit proves that any symmetry-projected thermal state of a local Hamiltonian without ordinary SSB necessarily exhibits SW-SSB~\cite{lessa2025,liu2025,weinstein2025}.

\paragraph*{Illustrations.---}
To make these general results concrete, we first test the geometric hierarchy of Eq.~\eqref{eq:chain} in two dephasing models. At finite distance, one model saturates the inequalities whereas the other keeps the diagnostics strictly separated, yet both yield the same asymptotic LRO criterion. We then turn to a solvable thermal model to benchmark Eq.~\eqref{eq:thermalbound}, showing how the mean injected energy $\Delta E_G$ captures state-dependent thermal response beyond the uniform operator-norm bound of Eq.~\eqref{eq:localbound}.

The saturation regime is realized in an open ZXZ chain [Fig.~\ref{fig:summary}(b)]~\cite{lu2026,pfeuty1970}. Starting from the ground state with both sublattice parities fixed to $+1$, we apply complete $\hat{X}$ dephasing on the even sublattice, decomposing the state into orthogonal conditional Ising blocks labeled by the even-site $\hat{X}$-eigenvalue syndromes $s$. For two odd-sublattice endpoints $n$ and $m$, the conditional branch correlations $c_s$ may vary in sign across sectors, but all share the same magnitude $|c_s|=|C_{nm}|$, where $C_{nm}$ is the corresponding Ising two-point correlation. This syndrome-independent magnitude yields $F_{\mathcal O}=|C_{nm}|$ and $D_s=R_1=F_{\mathcal O}^2=C_{nm}^2$, with $Q=4(1-C_{nm}^2)$, thereby saturating both inequalities surrounding $D_s$ in Eq.~\eqref{eq:chain} at all distances.

By contrast, the two-dimensional edge-dephasing construction  realizes the complementary regime in which the inequalities surrounding $D_s$ are generally strict. On an open square lattice of $N$ sites, we prepare $\rho_0=|+\rangle\langle+|^{\otimes N}$ and independently apply $\mathcal E_{\ell,p}(\rho)=(1-p)\rho+p\hat{Z}_i\hat{Z}_j\rho \hat{Z}_i\hat{Z}_j$ on each nearest-neighbor edge $\ell=\langle ij\rangle$, with $0\le p\le1/2$. The output state is diagonal in the syndrome configuration basis $\{|X\rangle\}$ with probabilities $P_X$, and the pair operation maps each syndrome $X$ to its endpoint-flipped partner $X^{ij}$. With zero-weight pairs omitted, these probabilities give
\begin{equation}
  D_s = \sum_X \frac{2P_X P_{X^{ij}}}{P_X + P_{X^{ij}}},\qquad
  F_\mathcal{O} = \sum_X \sqrt{P_X P_{X^{ij}}}.
  \label{eq:2dmeans}
\end{equation}
The common diagonal basis enforces $F_\mathcal O=R_1$ and reduces the hierarchy to $F_\mathcal O^2\le D_s\le F_\mathcal O$. Because $P_X=P_{X^{ij}}$ does not hold for all $X$ at $0<p<1/2$, the harmonic-mean sum $D_s$ and the geometric-mean sum $F_\mathcal O$ satisfy the strict ordering $F_\mathcal O^2 < D_s < F_\mathcal O$, placing the state strictly inside the allowed region [Fig.~\ref{fig:summary}(a)]. Together with the exact saturation in the ZXZ chain, this confirms the tightness and rigor of the two-sided bounds in Eq.~\eqref{eq:chain}. Under these rigorous bounds, our numerical results in Fig.~\ref{fig:summary}(c) resolve the transition across $p_c$, supporting the effectiveness of $D_s$ in diagnosing SW-SSB in the asymptotic limit of Eq.~\eqref{eq:lro}~\cite{supplement}.

Finally, to benchmark the thermal bounds in Eqs.~\eqref{eq:thermalbound} and \eqref{eq:localbound}, we consider a solvable parity-projected state of noninteracting spins with $\hat{H}=-h\sum_z\hat{X}_z$ and pair generator $\hat{G}=\hat{Z}_i\hat{Z}_j$~\cite{supplement}. In the thermodynamic limit at fixed $u=\beta h>0$, one obtains the separation-independent values $D_s=\sech(2u)$ and $\Delta E_G=4h\tanh u$, alongside the commutator norm $J_O=2h$. Substituting these into Eqs.~\eqref{eq:thermalbound} and \eqref{eq:localbound} orders the exact deficit, the injected-energy bound, and the commutator-norm bound as $\sech(2u)\ge e^{-4u\tanh u}\ge e^{-4u}$ [Fig.~\ref{fig:summary}(d)]. Both bounds guarantee a positive thermal deficit uniformly in separation, yet the injected-energy bound is strictly tighter because $\Delta E_G$ captures the actual thermal endpoint response $\tanh u$ rather than the state-independent worst-case scale $J_O$, directly validating our second main result in the thermal regime.

\paragraph*{Operational access.---}
While response spectroscopy determines the QFI in thermal equilibrium via fluctuation--dissipation relations~\cite{hauke2016}, general prepared states can be probed through a direct three-angle readout protocol. Combining the SLD variational functional~\cite{macieszczak2013,chabuda2020} for a Hermitian observable $\hat{M}$ with the two-eigenvalue parameter-shift identity for involutions~\cite{schuld2019} yields the lower bound
\begin{equation}
    J(\hat{M}) \equiv 2\left[\langle \hat{M}\rangle_{\pi/4} - \langle \hat{M}\rangle_{-\pi/4}\right] - \langle \hat{M}^2\rangle_0 \le Q,
    \label{eq:variationalreadout}
\end{equation}
where $\langle \hat{A}\rangle_\theta \equiv \Tr(\hat{A}\rho_\theta)$ requires measurements at only three rotation angles, $\theta=0,\pm\pi/4$. Supplementing $J(\hat{M})$ with an upper bound $\Delta_{\rm c}$ on the variational optimality gap,directly bounds the QFI via $Q\le J(\hat{M})+\Delta_{\rm c}$. Consequently, the condition $J(\hat{M})+\Delta_{\rm c}<4$ provides an operational certification of a nonzero deficit $D_s>0$~\cite{supplement}.

\paragraph*{Discussion.---}
The geometric framework developed here extends naturally beyond Hermitian involutions $\hat{G}^2 = \hat{I}$. Through a two-quadrature construction, we generalize the deficit hierarchy in Eq.~\eqref{eq:chain} to neutral unitary pairs formed from unitary charged endpoint operators~\cite{supplement}.In finite-dimensional systems with $U(1)$ symmetry, however, no unitary operator can carry a definite nonzero charge, so extending this construction to nonunitary charged endpoint operators requires an alternative formulation. On the operational side, developing sufficiently tight, measurement-accessible upper bounds $\Delta_{\rm c}$ on the variational optimality gap via controlled SLD approximations would provide a route to certifying $D_s>0$ in general prepared states through restricted measurements.

\begin{acknowledgments}
We thank Xilai Pang and Haoyue Qi for insightful discussions. This work is supported by NSFC (Grants No.GG2030007011 and No.GG2030040453) and Innovation Program for Quantum Science and Technology (No.2021ZD0302004).
\end{acknowledgments}

\onecolumngrid
\clearpage
\setcounter{equation}{0}
\setcounter{figure}{0}
\setcounter{table}{0}
\renewcommand{\theequation}{S\arabic{equation}}
\renewcommand{\thefigure}{S\arabic{figure}}
\renewcommand{\thetable}{S\arabic{table}}
\providecommand{\theHequation}{}
\providecommand{\theHfigure}{}
\providecommand{\theHtable}{}
\renewcommand{\theHequation}{supp.\arabic{equation}}
\renewcommand{\theHfigure}{supp.\arabic{figure}}
\renewcommand{\theHtable}{supp.\arabic{table}}
\pdfbookmark[0]{Supplemental Material}{supplement.title}
\begin{center}
{\large\bfseries Supplemental Material for ``Quantum Fisher Information Deficit as a Geometric Signature of Strong-to-Weak Spontaneous Symmetry Breaking''\par}
\vspace{0.8em}
Zhang-Chi Yin\textsuperscript{1} and Wei Zheng\textsuperscript{1,2,3,\href{mailto:zw8796@ustc.edu.cn}{*}}\par
\vspace{0.3em}
{\small\itshape
\textsuperscript{1}Hefei National Research Center for Physical Sciences at the Microscale and School of Physical Sciences, University of Science and Technology of China, Hefei 230026, China\par
\textsuperscript{2}CAS Center for Excellence in Quantum Information and Quantum Physics, University of Science and Technology of China, Hefei 230026, China\par
\textsuperscript{3}Hefei National Laboratory, University of Science and Technology of China, Hefei 230088, China\par
}
\end{center}
\vspace{0.5em}

We use the root (Uhlmann) fidelity $F(\rho,\sigma)=\Tr\sqrt{\sqrt\rho\,\sigma\sqrt\rho}$. The pair fidelity is $F_\mathcal O\equiv F(\rho,\hat{G}\rho \hat{G})$, where the neutral generator $\hat{G}$ is defined below. We write $Q[\rho;\hat{G}]$ for the standard SLD quantum Fisher information of the unitary family generated by $\hat{G}$ and $D_s\equiv1-Q/4$ when $\hat{G}$ is a Hermitian involution. Throughout, $k_B=\hbar=1$.

Section~S1 derives the hierarchy $F_\mathcal{O}^2\le D_s\le R_1\le F_\mathcal{O}$ and the long-range criterion with an explicit order of limits. Section~S2 proves the thermal bounds using relative entropy and energy-transfer spectra, and specifies the equilibrium-response conventions. Section~S3 extends the hierarchy to two quadratures when a unitary charged endpoint operator exists, a condition that fails for generic finite-dimensional $U(1)$ systems. Section~S4 derives exact three-angle access for a fixed readout and the SLD variational identity, then states the additional upper bound needed to certify $D_s>0$. Sections~S5 and S6 apply these results to the dephased ZXZ chain and the two-dimensional edge-dephasing model.

\section{S1. General deficit bounds}

Let $\rho=\sum_a\lambda_a|a\rangle\langle a|$ be a density matrix and $\hat{G}=\hat{G}^\dagger$ a Hermitian involution, $\hat{G}^2=\hat{I}$. As in the main text, we take $\hat{G}=\hat{O}_i\hat{O}_j$ to be the neutral pair of disjoint local charged Hermitian involutions $\hat{O}_i$ and $\hat{O}_j$. The orbit is $\rho_\theta=e^{-\ii\theta \hat{G}}\rho\,e^{\ii\theta \hat{G}}$, and $Q[\rho;\hat{G}]$ denotes its SLD quantum Fisher information at $\theta=0$. The SLD $\hat{\mathsf L}=\hat{\mathsf L}^\dagger$ satisfies the Lyapunov equation $-\ii[\hat{G},\rho]=(\rho\hat{\mathsf L}+\hat{\mathsf L}\rho)/2$. Its matrix elements in the eigenbasis of $\rho$ are
\begin{equation}
 \mathsf L_{ab}=\frac{2\ii(\lambda_a-\lambda_b)}{\lambda_a+\lambda_b}G_{ab}
 \label{eq:s-sld}
\end{equation}
for $\lambda_a+\lambda_b>0$, with zero-denominator terms set to zero. Substituting Eq.~\eqref{eq:s-sld} into $Q=\Tr(\rho\hat{\mathsf L}^2)$ and using $\Tr(\rho\hat{\mathsf L}^2)=\sum_{a,b}\lambda_a \mathsf L_{ab}\mathsf L_{ba}$ gives
\begin{equation}
 Q[\rho;\hat{G}]=2\sum_{a,b:\lambda_a+\lambda_b>0}
 \frac{(\lambda_a-\lambda_b)^2}{\lambda_a+\lambda_b}|G_{ab}|^2 .
 \label{eq:s-qfi}
\end{equation}
Terms with $\lambda_a=\lambda_b$ vanish, while support--kernel transitions remain in the sum. To obtain the deficit, use $\hat{G}^2=\hat{I}$ and $\Tr(\rho \hat{G}^2)=1$ to write
\begin{equation}
 1=\Tr(\rho \hat{G}^2)=\sum_{a,b:\lambda_a+\lambda_b>0}\lambda_a|G_{ab}|^2
 =\sum_{a,b:\lambda_a+\lambda_b>0}\frac{\lambda_a+\lambda_b}{2}|G_{ab}|^2 ,
 \label{eq:s-normalization}
\end{equation}
where the last equality symmetrizes $a\leftrightarrow b$. Subtracting one quarter of Eq.~\eqref{eq:s-qfi} from Eq.~\eqref{eq:s-normalization}, with $(\lambda_a+\lambda_b)^2-(\lambda_a-\lambda_b)^2=4\lambda_a\lambda_b$, gives the harmonic-mean representation
\begin{equation}
 \Ds=1-\frac{Q}{4}
 =\sum_{a,b:\lambda_a+\lambda_b>0}
 \frac{2\lambda_a\lambda_b}{\lambda_a+\lambda_b}|G_{ab}|^2 .
 \label{eq:s-deficit}
\end{equation}
The R\'enyi-1 (Wightman) correlator for the same generator has the spectral representation
\begin{equation}
 \Rone=\Tr(\sqrt\rho\,\hat{G}\sqrt\rho\,\hat{G})
 =\sum_{a,b}\sqrt{\lambda_a\lambda_b}\,|G_{ab}|^2 .
 \label{eq:s-r1}
\end{equation}
Comparing Eqs.~\eqref{eq:s-deficit} and \eqref{eq:s-r1}, apply $2xy/(x+y)\le\sqrt{xy}$ for $x,y\ge0$, including $xy=0$, to each pair $(a,b)$ to obtain
\begin{equation}
 \Ds\le \Rone .
 \label{eq:s-dle-r1}
\end{equation}
In terms of the Wigner--Yanase information, $I_{\rm WY}(\rho,\hat{G})=-\tfrac12\Tr[\sqrt\rho,\hat{G}]^2=1-\Rone$, this is the standard metric inequality $I_{\rm WY}\le Q/4$~\cite{luo2003,luo2004,gibiliscoimparato2009,gibilisco2008}. The harmonic and geometric means weight the same squared matrix elements, so the comparison holds at finite size.

For the upper bound $\Rone\le F_\mathcal{O}$, write the fidelity between $\rho$ and $\hat{G}\rho \hat{G}$ as
\begin{equation}
 F(\rho,\hat{G}\rho \hat{G})=\|\sqrt\rho\,\hat{G}\sqrt\rho\|_1
 =\max_{\hat{V}\ \text{unitary}}\bigl|\Tr\bigl(\hat{V}\sqrt\rho\,\hat{G}\sqrt\rho\bigr)\bigr| ,
 \label{eq:s-tracenorm}
\end{equation}
using trace-norm duality, $\|\hat{A}\|_1=\max_{\hat{V}^\dagger \hat{V}=\hat{I}}|\Tr(\hat{V}\hat{A})|$, with the maximizing unitary chosen from the polar decomposition of $\hat{A}=\sqrt\rho\,\hat{G}\sqrt\rho$. Choosing $\hat{V}=\hat{G}$ and cycling the trace gives
\begin{equation}
 \Tr\!\left(\hat{G}\sqrt\rho\,\hat{G}\sqrt\rho\right)=\Tr\!\left(\sqrt\rho\,\hat{G}\sqrt\rho\,\hat{G}\right)=\Rone .
 \label{eq:s-r1-trace}
\end{equation}
The trace is real and nonnegative by Eq.~\eqref{eq:s-r1}. With the choice $\hat{V}=\hat{G}$ in Eq.~\eqref{eq:s-tracenorm}, it therefore gives
\begin{equation}
 \Rone\le F(\rho,\hat{G}\rho \hat{G})=\FO .
 \label{eq:s-r1-le-f}
\end{equation}
This is the affinity--root-fidelity bound~\cite{audenaert2007}.

The lower bound $F_\mathcal{O}^2\le\Ds$ relates the finite overlap to the infinitesimal metric through the convex-roof structure of QFI. For a pure-state ensemble $\mathcal A=\{w_k,|\psi_k\rangle\}$ realizing $\rho=\sum_k w_k|\psi_k\rangle\langle\psi_k|$, define $g_k=\langle\psi_k|\hat{G}|\psi_k\rangle$. The convex-roof theorem~\cite{yu2013}, conjectured in Ref.~\cite{tothpetz2013}, gives
\begin{equation}
 \frac14 Q[\rho;\hat{G}]=\min_{\mathcal A}\sum_k w_k\bigl(1-g_k^2\bigr),
 \qquad
 \Ds=\max_{\mathcal A}\sum_k w_k g_k^2 ,
 \label{eq:s-defensemble}
\end{equation}
where the second equality uses $\sum_k w_k=1$. To express the fidelity in ensemble form, set $\hat{B}=\sqrt\rho\,\hat{G}\sqrt\rho$ on $\operatorname{supp}\rho$. The Hughston--Jozsa--Wootters (HJW) parametrization~\cite{hjw1993} realizes every ensemble of $\rho$ through vectors $|u_k\rangle$ satisfying
\begin{equation}
 \sqrt{w_k}\,|\psi_k\rangle=\sqrt\rho\,|u_k\rangle,
 \qquad
 \sum_k|u_k\rangle\langle u_k|=\hat{\Pi}_{\operatorname{supp}\rho} .
 \label{eq:s-hjw}
\end{equation}
Taking the expectation of $\hat{G}$ in $|\psi_k\rangle$ and using Eq.~\eqref{eq:s-hjw} gives $w_k g_k=\langle u_k|\hat{B}|u_k\rangle$. The optimized mean absolute branch expectation is therefore
\begin{equation}
 \max_{\mathcal A}\sum_k w_k|g_k|
 =\max_{\{u_k\}}\sum_k\bigl|\langle u_k|\hat{B}|u_k\rangle\bigr|
 =\Tr|\hat{B}|=\FO .
 \label{eq:s-fensemble}
\end{equation}
To obtain the second equality, use $|\langle u_k|\hat{B}|u_k\rangle|\le\langle u_k|\,|\hat{B}|\,|u_k\rangle$ to bound $\sum_k|\langle u_k|\hat{B}|u_k\rangle|\le\Tr|\hat{B}|$ for every resolution of the identity. Choosing $\{|u_k\rangle\}$ as an eigenbasis of the Hermitian operator $\hat{B}$ attains the bound. For any fixed ensemble, Cauchy--Schwarz also gives
\begin{equation}
 \Bigl(\sum_k w_k|g_k|\Bigr)^2\le\Bigl(\sum_k w_k\Bigr)\Bigl(\sum_k w_k g_k^2\Bigr)=\sum_k w_k g_k^2 .
 \label{eq:s-cs}
\end{equation}
Evaluate Eq.~\eqref{eq:s-cs} on a fidelity-maximizing ensemble in Eq.~\eqref{eq:s-fensemble}, then bound the right-hand side by the ensemble maximum in Eq.~\eqref{eq:s-defensemble}. This yields
\begin{equation}
 \FO^2\le\Ds .
 \label{eq:s-f2-le-d}
\end{equation}
The two optimizers need not coincide, and $\rho$ may be degenerate. The HJW resolution handles zero eigenvalues on $\operatorname{supp}\rho$. Evaluating Cauchy--Schwarz on a fidelity-optimal ensemble requires no assumption that it also maximizes the deficit, so the result is an inequality rather than an identity.

Combining Eqs.~\eqref{eq:s-dle-r1}, \eqref{eq:s-r1-le-f}, and \eqref{eq:s-f2-le-d} gives the main-text hierarchy,
\begin{equation}
 \FO^2\le\Ds\le\Rone\le\FO .
 \label{eq:s-chain}
\end{equation}
A complementary spectral check of Eq.~\eqref{eq:s-chain} follows by writing the spectral quantities as weighted means. Define $w_{ab}=(\lambda_a+\lambda_b)|G_{ab}|^2/2$, normalized by $\sum_{ab}w_{ab}=1$ from Eq.~\eqref{eq:s-normalization}, and $a_{ab}=2\sqrt{\lambda_a\lambda_b}/(\lambda_a+\lambda_b)\in[0,1]$ for $\lambda_a+\lambda_b>0$. When $\lambda_a+\lambda_b=0$, the weight vanishes and we set $a_{ab}=0$. Then the two quantities are the weighted means of $a_{ab}$ and $a_{ab}^2$,
\begin{equation}
 \Rone=\sum_{ab}w_{ab}\,a_{ab},
 \qquad
 \Ds=\sum_{ab}w_{ab}\,a_{ab}^2 ,
 \label{eq:s-meanform}
\end{equation}
where $a_{ab}^2\le a_{ab}$ gives $\Ds\le\Rone$, while Jensen's inequality $(\sum w a)^2\le\sum w a^2$ gives $\Rone^2\le\Ds$. The latter is weaker than Eq.~\eqref{eq:s-f2-le-d}: it uses only the spectral data, without the freedom of the fidelity-optimal ensemble. Two limiting cases illustrate the hierarchy. For a pure state $\rho=|\psi\rangle\langle\psi|$, let $g=\langle\psi|\hat{G}|\psi\rangle$. Then $F_\mathcal{O}=|g|$ and $D_s=R_1=g^2$, which saturate the lower link, $F_\mathcal{O}^2=D_s=R_1\le F_\mathcal{O}$. If $[\hat{G},\rho]=0$, the orbit is constant, with $Q=0$ and $D_s=R_1=F_\mathcal{O}=1$.

To apply Eq.~\eqref{eq:s-chain} to spatial order, let $X^{(L)}(r)$ be a diagnostic at separation $r=d(i,j)$. As in the main text, take the thermodynamic limit at fixed $r$, $X(r)=\lim_{L\to\infty}X^{(L)}(r)$, followed by the long-distance limit
\begin{equation}
 \mathcal L[X]\equiv\lim_{r\to\infty}X(r).
 \label{eq:s-liminf}
\end{equation}
Since Eq.~\eqref{eq:s-chain} holds at every $L$ and $r$, these limits give
\begin{equation}
 \mathcal L[\FO]>0
 \Longleftrightarrow
 \mathcal L[\Ds]>0
 \Longleftrightarrow
 \mathcal L[\Rone]>0 .
 \label{eq:s-lro}
\end{equation}
The successive limits must exist; a finite-size plateau alone does not establish them. If the ordinary charged two-point correlator vanishes in the same channel and the state remains in a fixed strong-symmetry sector, Eq.~\eqref{eq:s-lro} gives the SW-SSB criterion of the main text. The three diagnostics therefore have equivalent LRO criteria, although their finite-distance values, correlation lengths, and critical exponents can differ.

If algebraic decay is independently established, $\FO(r)\sim r^{-\eta_F}$ and $\Ds(r)\sim r^{-\eta_D}$, the bounds $F_\mathcal{O}^2\le D_s\le F_\mathcal{O}$ imply
\begin{equation}
 \eta_F\le\eta_D\le2\eta_F .
 \label{eq:s-exponents}
\end{equation}
An analogous statement for a R\'enyi-2 correlator requires additional model-specific structure~\cite{lessa2025,weinstein2025}.

\section{S2. Thermal proofs and response conventions}

The thermal bound follows by combining relative-entropy bounds on nonlinear overlaps~\cite{zhanglocal2026} with the hierarchy in S1. Let $\hat{P}$ be a nonzero orthogonal projector satisfying $[\hat{H},\hat{P}]=0$, and define
\begin{equation}
 \rho_{\beta,P}=\frac{\hat{P}e^{-\beta \hat{H}}}{Z_P},
 \qquad
 Z_P=\Tr(\hat{P}e^{-\beta \hat{H}}),
 \qquad
 0<\beta<\infty .
 \label{eq:s-sector-gibbs}
\end{equation}
Assume $\hat{G}=\hat{G}^\dagger$, $\hat{G}^2=\hat{I}$, and $[\hat{G},\hat{P}]=0$. An invariant support suffices for the thermal inequality. Its strong-symmetry interpretation further requires $\hat{P}$ to select a fixed charge sector, such as a $\mathbb Z_2$ parity sector.

The mean energy injected by the pair operation is
\begin{equation}
 \Delta E_G\equiv
 \Tr\!\left[\rho_{\beta,P}(\hat{G}\hat{H}\hat{G}-\hat{H})\right].
 \label{eq:s-dEdef}
\end{equation}
An individual transition $E_m-E_n$ can have either sign. Thermal passivity gives a nonnegative \emph{average} $\Delta E_G$, as follows directly from relative entropy. The state $\sigma=\hat{G}\rho_{\beta,P}\hat{G}$ has the same support as the original thermal state, and
\begin{align}
 S(\sigma\Vert\rho_{\beta,P})
 &=\Tr\!\left[\sigma(\log\sigma-\log\rho_{\beta,P})\right]\nonumber\\
 &=\beta\,\Tr\!\left[\rho_{\beta,P}(\hat{G}\hat{H}\hat{G}-\hat{H})\right]
 =\beta\Delta E_G\ge0 .
 \label{eq:s-relative}
\end{align}
The second equality uses $\log(\hat{G}\rho \hat{G})=\hat{G}(\log\rho)\hat{G}$ on $\operatorname{Ran}\hat{P}$ and $[\hat{G},\hat{P}]=0$.

Monotonicity of the Petz R\'enyi divergences in their order~\cite{mosonyi2011} gives
\begin{equation}
 D^{\rm P}_{1/2}(\sigma\Vert\rho_{\beta,P})
 \le
 D^{\rm P}_{1}(\sigma\Vert\rho_{\beta,P})
 =
 S(\sigma\Vert\rho_{\beta,P}) ,
 \label{eq:s-petz-order}
\end{equation}
with
\begin{equation}
 D^{\rm P}_{1/2}(\sigma\Vert\rho_{\beta,P})
 =-2\log \Tr(\sqrt{\sigma}\sqrt{\rho_{\beta,P}})
 =-2\log R_1 .
 \label{eq:s-petz}
\end{equation}
The overlap in Eq.~\eqref{eq:s-petz} is the R\'enyi-1/Wightman correlator $R_1$, distinct from the Uhlmann fidelity. Equations~\eqref{eq:s-relative}--\eqref{eq:s-petz} give
\begin{equation}
 R_1\ge e^{-\beta\Delta E_G/2}.
 \label{eq:s-r1thermal}
\end{equation}
The bound $R_1\le F_\mathcal O$ in Eq.~\eqref{eq:s-r1-le-f} also gives
\begin{equation}
 F_\mathcal O\ge R_1\ge e^{-\beta\Delta E_G/2}.
 \label{eq:s-fthermal}
\end{equation}
Combining $F_\mathcal O^2\le D_s$ with this estimate, or using $R_1^2\le D_s$ from Eq.~\eqref{eq:s-meanform}, yields the same thermal lower bound. The two inequalities are not equivalent for arbitrary states: $F_\mathcal O\ge e^{-\beta\Delta E_G/2}$ alone gives no corresponding lower bound on $R_1$. Thus
\begin{equation}
 D_s(\rho_{\beta,P},\hat{G})\ge e^{-\beta\Delta E_G}.
 \label{eq:s-thermalbound}
\end{equation}
A finite mean injected energy therefore guarantees a positive deficit.

Locality converts this state-dependent result into a uniform bound. Since $\hat{G}\hat{H}\hat{G}-\hat{H}=\hat{G}[\hat{H},\hat{G}]$,
\begin{equation}
 0\le\Delta E_G
 \le\|\hat{G}\hat{H}\hat{G}-\hat{H}\|
 =\|[\hat{H},\hat{G}]\|.
 \label{eq:s-commnorm}
\end{equation}
For a neutral pair $\hat{G}=\hat{O}_i\hat{O}_j$ of disjoint commuting Hermitian involutions,
\begin{equation}
 \|[\hat{H},\hat{G}]\|
 \le \|[\hat{H},\hat{O}_i]\|+\|[\hat{H},\hat{O}_j]\|
 \le2J_O,
 \qquad
 J_O\equiv\sup_{L,\hat{O}\in\mathcal O}\|[\hat{H}_L,\hat{O}]\|.
 \label{eq:s-localcomm}
\end{equation}
The family $\mathcal O$ contains all site copies of the chosen endpoint operator $\hat{O}$. Finite-range bounded interactions give $J_O<\infty$, as do uniformly summable interactions with a volume-independent endpoint energy scale. Combining the bounds gives
\begin{equation}
 D_s^{(L)}(i,j)
 \ge e^{-\beta\Delta E_G}
 \ge e^{-\beta\|[\hat{H},\hat{G}]\|}
 \ge e^{-2\beta J_O}>0 .
 \label{eq:s-localfloor}
\end{equation}
The final lower bound is independent of system size and separation, so it remains positive in the long-distance limit at any fixed finite temperature. Thermal SW-SSB additionally requires the absence of ordinary charged LRO in the same channel.

The energy-transfer representation gives a second route to a local bound. Within $\operatorname{Ran}\hat{P}$, write $\hat{H}|n\rangle=E_n|n\rangle$ and $\lambda_n=e^{-\beta E_n}/Z_P$, and define the normalized measure
\begin{equation}
 d\mu_G(\omega)=\sum_{mn}\lambda_n|G_{mn}|^2\,
 \delta\bigl[\omega-(E_m-E_n)\bigr]\,d\omega ,
 \qquad \int d\mu_G(\omega)=1 ,
 \label{eq:s-mu}
\end{equation}
where normalization follows by tracing $\hat{G}^2=\hat{I}$ against $\rho_{\beta,P}$. Pairing $(m,n)$ with its reverse $(n,m)$ and using detailed balance, $\lambda_m=e^{-\beta(E_m-E_n)}\lambda_n$, expresses the QFI as an integral over this measure. The deficit entering Eq.~\eqref{eq:s-chain} has the complementary kernel,
\begin{equation}
 \frac{Q}{4}=\int\tanh^2\!\Bigl(\frac{\beta\omega}{2}\Bigr)d\mu_G(\omega),
 \qquad
 \Ds=\int\sech^2\!\Bigl(\frac{\beta\omega}{2}\Bigr)d\mu_G(\omega) .
 \label{eq:s-kernels}
\end{equation}
Equation~\eqref{eq:s-kernels} separates the response by energy scale. The QFI kernel approaches unity for $|\omega|\gg T$, while the deficit kernel is largest near $\omega=0$. A sequence with $D_s\to0$ must therefore shift its pair-induced spectral weight to increasingly high frequencies. A uniform endpoint energy scale constrains this weight through the second moment
\begin{equation}
 \int\omega^2\,d\mu_G(\omega)
 =\Tr\bigl\{\rho_{\beta,P}\,[\hat{H},\hat{G}]^\dagger[\hat{H},\hat{G}]\bigr\}\le\|[\hat{H},\hat{G}]\|^2 ,
 \label{eq:s-secondmoment}
\end{equation}
obtained from the definition of $d\mu_G$ using $(E_m-E_n)^2|G_{mn}|^2=|[\hat{H},\hat{G}]_{mn}|^2$ and the norm bound on $[\hat{H},\hat{G}]$. The inequality $\sech^2 x\ge e^{-2|x|}$, Jensen's inequality for the exponential, and Cauchy--Schwarz for the first absolute moment then give
\begin{equation}
 \Ds\ge\int e^{-\beta|\omega|}\,d\mu_G
 \ge e^{-\beta\int|\omega|\,d\mu_G}
 \ge e^{-\beta\|[\hat{H},\hat{G}]\|} ,
 \label{eq:s-normproof}
\end{equation}
where the last step uses $\int|\omega|\,d\mu_G\le\bigl(\int\omega^2 d\mu_G\bigr)^{1/2}\le\|[\hat{H},\hat{G}]\|$ from Eq.~\eqref{eq:s-secondmoment}. This independent norm-level bound is generally weaker than Eq.~\eqref{eq:s-thermalbound}: it replaces the signed mean $\Delta E_G$ by an operator norm that can be substantially larger for near-commuting generators.

Fixed-parity independent spins illustrate the difference between these energy scales. Take $\hat{H}=-h\sum_{z=1}^L\hat{X}_z$, with parity $\hat{U}=\prod_z\hat{X}_z$ and projector $\hat{P}_q=(\hat{I}+q\hat{U})/2$ for $q=\pm1$. For $\hat{G}=\hat{Z}_i\hat{Z}_j$ at distinct sites $i\ne j$, conjugation gives $\hat{G}\hat{H}\hat{G}-\hat{H}=2h(\hat{X}_i+\hat{X}_j)$. The injected energy is thus the sum of two single-site responses. The partition function is
\begin{equation}
 Z_q=\frac12\bigl[(2\cosh u)^L+q(2\sinh u)^L\bigr],
 \qquad u=\beta h ,
 \label{eq:s-spinZ}
\end{equation}
and at fixed $u>0$, the parity correction in Eq.~\eqref{eq:s-spinZ} vanishes exponentially with $L$ as $L\to\infty$. Evaluating Eq.~\eqref{eq:s-dEdef} in this limit gives
\begin{equation}
 \Delta E_G^{\infty}=4h\tanh u ,
 \qquad
 \Ds^{\infty}=\sech(2u) ,
 \qquad
 \Rone^{\infty}=\FO^{\infty}=\sech^2 u ,
 \label{eq:s-spinexact}
\end{equation}
and hence $\sech(2u)\ge e^{-4u\tanh u}\ge e^{-4u}$ for every $u>0$, with the injected-energy bound tighter than the norm bound. The analytic curve extends continuously to $u=0$, where both bounds equal one. In main-text Fig.~2(d), $\Delta E$ denotes $\Delta E_G$.

A factorization of the separated-pair spectrum into a convolution of one-endpoint spectra does not follow from Eq.~\eqref{eq:s-mu} alone. It would require a thermodynamic limit of the nonsymmetrized spectral measures and a clustering or Lieb--Robinson estimate for connected dynamical correlations. These assumptions are unnecessary for Eq.~\eqref{eq:s-localcomm}. Fourier transforming Eq.~\eqref{eq:s-kernels} also gives the exact time-domain representation
\begin{equation}
 \Ds=\int_{-\infty}^{\infty}dt\,K_\beta(t)\,\Tr\bigl[\rho_{\beta,P}\,\hat{G}(t)\hat{G}\bigr],
 \qquad
 K_\beta(t)=\frac{2t/\beta^2}{\sinh(\pi t/\beta)} ,
 \label{eq:s-timekernel}
\end{equation}
where continuity fixes the value at $t=0$. This identity alone does not control an approximation of the full pair QFI by a reduced-state or finite-light-cone QFI. Local and marginal fidelity formulations of SW-SSB are available~\cite{divi2026,zhanglocal2026,liulocal2026}; the corresponding controlled approximation for the present QFI response remains open.

For equilibrium spectroscopy, define the retarded response $\chi_G(t)=\ii\Theta(t)\Tr\{\rho_{\beta,P}[\hat{G}(t),\hat{G}(0)]\}$. With the Fourier convention of Ref.~\cite{hauke2016}, the fluctuation--dissipation relation is
\begin{equation}
 Q[\rho_{\beta,P};\hat{G}]=\frac{4}{\pi}\int_0^\infty d\omega\,
 \tanh\!\Bigl(\frac{\beta\omega}{2}\Bigr)\chi_G''(\omega) .
 \label{eq:s-response}
\end{equation}
Elastic zero-frequency weight contributes to the deficit in Eq.~\eqref{eq:s-kernels} but vanishes from the response integral. A finite-band measurement therefore gives only a QFI lower bound. Certifying a positive deficit also requires an upper bound on the omitted high-frequency tail and control of systematic errors, as in the readout setting of Section~S4.

\section{S3. Conditional extension to unitary charged pairs}

The main-text $\mathbb Z_2$ construction uses Hermitian charged endpoints and an involutory neutral pair. A two-quadrature extension is possible if the charged endpoint operator is unitary. For continuous symmetries, the extension covers only settings that admit such an operator.

Let $\hat{W}_{ij}=\hat{O}_i\hat{O}_j^\dagger$ be a neutral unitary pair, $\hat{W}^\dagger \hat{W}=\hat{W}\hat{W}^\dagger=\hat{I}$. Suppressing the spatial indices, define its Hermitian quadratures
\begin{equation}
 \hat{C}=\frac{\hat{W}+\hat{W}^\dagger}{2},
 \qquad
 \hat{S}=\frac{\hat{W}-\hat{W}^\dagger}{2\ii}.
 \label{eq:S3_CS_def}
\end{equation}
Unitarity implies
\begin{equation}
 [\hat{C},\hat{S}]=0,
 \qquad
 \hat{C}^2+\hat{S}^2=\hat{I} .
 \label{eq:S3_circle_identity}
\end{equation}
These identities motivate the two-quadrature deficit
\begin{equation}
 D_s^{(U)}\equiv
 1-\frac14\left[Q(\rho;\hat{C})+Q(\rho;\hat{S})\right].
 \label{eq:S3_DQ_def}
\end{equation}
Because $Q(\rho;\hat{A})\le4\Tr(\rho \hat{A}^2)$ for every Hermitian $\hat{A}$, Eq.~\eqref{eq:S3_circle_identity} gives $0\le D_s^{(U)}\le1$.

In the eigenbasis $\rho=\sum_a\lambda_a|a\rangle\langle a|$,
\begin{equation}
 |C_{ab}|^2+|S_{ab}|^2
 =\frac12\left(|W_{ab}|^2+|W_{ba}|^2\right).
 \label{eq:S3_CS_elements}
\end{equation}
The QFI coefficient is symmetric under $a\leftrightarrow b$. Using this symmetry and unitarity in Eq.~\eqref{eq:S3_DQ_def} gives
\begin{equation}
 D_s^{(U)}
 =
 \sum_{a,b:\lambda_a+\lambda_b>0}
 \frac{2\lambda_a\lambda_b}{\lambda_a+\lambda_b}|W_{ab}|^2 .
 \label{eq:S3_DQ_spectral}
\end{equation}
Define
\begin{equation}
 R_1^{(U)}\equiv
 \Tr(\sqrt\rho\,\hat{W}\sqrt\rho\,\hat{W}^\dagger)
 =
 \sum_{a,b}\sqrt{\lambda_a\lambda_b}|W_{ab}|^2
\end{equation}
and
\begin{equation}
 F_W\equiv F(\rho,\hat{W}\rho \hat{W}^\dagger)
 =\|\sqrt\rho\,\hat{W}\sqrt\rho\|_1.
\end{equation}
The harmonic--geometric mean inequality gives $D_s^{(U)}\le R_1^{(U)}$. Trace-norm duality with the unitary $\hat{W}^\dagger$ gives $R_1^{(U)}\le F_W$. For the lower link, choose $\hat{V}$ to attain
$\|\sqrt\rho \hat{W}\sqrt\rho\|_1=|\Tr(\hat{V}\sqrt\rho \hat{W}\sqrt\rho)|$ and apply weighted Cauchy--Schwarz:
\begin{align}
 F_W^2
 &\le
 \left[
 \sum_{a,b:\lambda_a+\lambda_b>0}\frac{2\lambda_a\lambda_b}{\lambda_a+\lambda_b}|W_{ab}|^2
 \right]
 \left[
 \sum_{a,b}\frac{\lambda_a+\lambda_b}{2}|V_{ba}|^2
 \right]\nonumber\\
 &=D_s^{(U)},
 \label{eq:S3_lower_bound}
\end{align}
where the second bracket equals one by unitarity of $\hat{V}$. Together,
\begin{equation}
 F_W^2\le D_s^{(U)}\le R_1^{(U)}\le F_W .
 \label{eq:S3_main_bound}
\end{equation}
The involutory hierarchy is recovered when $\hat{W}=\hat{W}^\dagger$.

The extension requires a \emph{unitary} local charged operator. Suppose a finite-dimensional local Hilbert space has a continuous $U(1)$ charge operator $\hat N$ and a unitary $\hat{O}$ satisfying
\begin{equation}
 [\hat N,\hat{O}]=q\hat{O},
 \qquad q\ne0.
\end{equation}
Multiplying by $\hat{O}^\dagger$ from the left gives
\begin{equation}
 \hat{O}^\dagger\hat N \hat{O}-\hat N=q\hat{I}.
\end{equation}
Taking the trace gives $0=q\,\dim\mathcal H$, a contradiction. A definite nonzero-charge unitary therefore cannot exist in a finite-dimensional local $U(1)$ representation. Equation~\eqref{eq:S3_main_bound} applies conditionally, for example to rotor-like systems with a bounded unitary charged operator and controlled trace properties. It does not give a general finite-dimensional $U(1)$ construction. Ladder operators such as $\hat{S}^+$ and $\hat{a}^\dagger$ are nonunitary and require a different normalization and deficit definition. The fixed three-angle rule of Section~S4 also requires a separate analysis for $\hat{C}$ and $\hat{S}$, which are generally not involutions.

\section{S4. Exact three-angle access and QFI certification}

An involutory encoding $\hat{G}^2=\hat{I}$ gives exact three-angle access to the classical Fisher information of any fixed regular POVM. An SLD variational functional separately gives a QFI lower bound. Since $D_s=1-Q/4$, a lower bound on $Q$ is an upper bound on $D_s$; certifying a positive deficit requires an independent QFI upper bound.

Let $\mathsf E=\{\hat{E}_x\}$ be a POVM independent of $\theta$, with outcome probabilities
\begin{equation}
 p_x(\theta)=\Tr(\hat{E}_x\rho_\theta),
 \qquad
 \rho_\theta=e^{-\ii\theta \hat{G}}\rho e^{\ii\theta \hat{G}}.
\end{equation}
For $\hat{G}^2=\hat{I}$,
\begin{equation}
 e^{-\ii\theta \hat{G}}=\cos\theta\,\hat{I}-\ii\sin\theta\,\hat{G},
\end{equation}
so each outcome probability is harmonic,
\begin{equation}
 p_x(\theta)=a_x+b_x\cos2\theta+c_x\sin2\theta .
\end{equation}
The two-eigenvalue parameter-shift rule~\cite{schuld2019} gives
\begin{equation}
 \dot p_x(0)
 =
 p_x(\pi/4)-p_x(-\pi/4).
 \label{eq:s-shift-prob}
\end{equation}
At a regular working point with $p_x(0)>0$, the classical Fisher information is
\begin{equation}
 F_C[\mathsf E]
 =
 \sum_x
 \frac{\left[p_x(\pi/4)-p_x(-\pi/4)\right]^2}{p_x(0)}
 \le Q[\rho;\hat{G}].
 \label{eq:s-threeangle-cfi}
\end{equation}
Three settings suffice because the shift identity is exact. Estimating each probability still requires repeated preparations.

For $p_x(0)=0$, positivity requires $\dot p_x(0)=0$, while the limiting Fisher contribution can depend on the second-order behavior of $p_x(\theta)$. These outcomes require the continuous Fisher-information limit or a regularized working point. A zero count in finite data does not establish that the true probability vanishes.

Assigning real scores $m_x$ to the outcomes of this POVM defines
\begin{equation}
 J(\mathsf E,m)
 =
 2\sum_xm_x\dot p_x(0)
 -
 \sum_xp_x(0)m_x^2 .
 \label{eq:s-classical-score}
\end{equation}
Completing the square for each outcome gives
\begin{equation}
 F_C[\mathsf E]-J(\mathsf E,m)
 =
 \sum_{x:p_x(0)>0}
 p_x(0)
 \left[
 m_x-\frac{\dot p_x(0)}{p_x(0)}
 \right]^2
 \ge0,
 \label{eq:s-classical-gap}
\end{equation}
so the optimal score is $m_x^\star=\dot p_x/p_x$, with $\max_mJ(\mathsf E,m)=F_C[\mathsf E]$. Optimizing over measurements then gives $\sup_{\mathsf E}F_C[\mathsf E]=Q$.

The SLD variational formulation~\cite{macieszczak2013,chabuda2020} avoids reconstructing the outcome distribution. For a Hermitian readout $\hat{M}=\hat{M}^\dagger$, let
\begin{equation}
 \dot\rho=-\ii[\hat{G},\rho]
 =\frac12(\rho\hat{\mathsf L}+\hat{\mathsf L}\rho)
\end{equation}
define the SLD $\hat{\mathsf L}$. The variational functional is
\begin{equation}
 J(\hat{M})
 =
 2\Tr(\dot\rho \hat{M})-\Tr(\rho \hat{M}^2).
 \label{eq:s-score}
\end{equation}
Substituting the SLD equation and completing the square gives
\begin{equation}
 Q-J(\hat{M})
 =
 \Tr\!\left[\rho(\hat{M}-\hat{\mathsf L})^2\right]
 \ge0.
 \label{eq:s-var-gap}
\end{equation}
Consequently,
\begin{equation}
 Q=\max_{\hat{M}=\hat{M}^\dagger}J(\hat{M}),
 \qquad \hat{M}_{\rm opt}=\hat{\mathsf L}
 \label{eq:s-var}
\end{equation}
on the support of $\rho$.

With $\langle \hat{A}\rangle_\theta=\Tr(\hat{A}\rho_\theta)$ for an observable $\hat{A}$, the involutory shift rule evaluates the derivative entering $J(\hat{M})$ exactly:
\begin{equation}
 \partial_\theta\langle \hat{M}\rangle_{\theta=0}
 =
 \langle \hat{M}\rangle_{\pi/4}-\langle \hat{M}\rangle_{-\pi/4}.
\end{equation}
Hence
\begin{equation}
 J(\hat{M})
 =
 2\left[
 \langle \hat{M}\rangle_{\pi/4}
 -
 \langle \hat{M}\rangle_{-\pi/4}
 \right]
 -
 \langle \hat{M}^2\rangle_0 .
 \label{eq:s-threeangle-score}
\end{equation}
This is the main-text three-angle variational readout. Its cost depends on the structure of $\hat{M}$, the measurable decompositions of $\hat{M}$ and $\hat{M}^2$, and the effort needed to learn or construct the readout. Three-angle access alone does not imply efficiency independent of system size.

For any fixed measurement or variational readout,
\begin{equation}
 F_C\le Q,
 \qquad
 J(\hat{M})\le Q.
\end{equation}
Therefore
\begin{equation}
 D_s=1-\frac{Q}{4}
 \le
 1-\frac{F_C}{4},
 \qquad
 D_s\le1-\frac{J(\hat{M})}4.
 \label{eq:s-deficit-upper}
\end{equation}
These are \emph{upper} bounds on the deficit, so a measured Fisher information alone cannot certify $D_s>0$.

To certify a positive deficit, suppose an independent argument bounds the variational optimality gap, using the main-text notation $\Delta_{\rm c}$:
\begin{equation}
 0\le Q-J(\hat{M})\le\Delta_{\rm c},
 \label{eq:s-gap-cert}
\end{equation}
and independent validation data give $J(\hat{M})\in[J_-,J_+]$ at the stated confidence level. Then
\begin{equation}
 Q\in
 \left[
 \max(0,J_-),
 \min(4,J_++\Delta_{\rm c})
 \right]
\end{equation}
and hence
\begin{equation}
 \max\!\left(
 0,1-\frac{J_++\Delta_{\rm c}}4
 \right)
 \le D_s\le
 \min\!\left(
 1,1-\frac{\max(0,J_-)}4
 \right).
 \label{eq:s-certifiedinterval}
\end{equation}
A positive lower endpoint certifies the deficit. The upper bound in Eq.~\eqref{eq:s-gap-cert} may follow from an analytically optimal readout, a controlled SLD approximation, or an independent QFI upper bound.

For an approximate SLD, define the equation residual
\begin{equation}
 \hat{\mathcal R}
 =
 \dot\rho-\frac12(\rho \hat{M}+\hat{M}\rho),
\end{equation}
so that, in the eigenbasis of $\rho$,
\begin{equation}
 Q-J(\hat{M})
 =
 2\sum_{a,b:\lambda_a+\lambda_b>0}
 \frac{|\mathcal R_{ab}|^2}{\lambda_a+\lambda_b}.
 \label{eq:s-residual-gap}
\end{equation}
Small eigenvalues of $\rho$ amplify residual errors in this weighted sum, so a small unweighted residual norm alone does not control the gap.

The protocol requires reproducible preparation of the target state, a verified symmetry sector, and control of the pair rotation. The variational readout should be trained on data separate from those used to report $J$ or $F_C$. Generic efficient protocols for state-agnostic detection of arbitrary unknown strongly symmetric states face computational obstructions~\cite{feng2025}. Those protocols concern a different access setting.

\section{S5. Dephased ZXZ chain}

The open ZXZ chain saturates both hierarchy bounds adjacent to the deficit. For even length $L$, take
\begin{equation}
 \hat{H}_{\rm ZXZ}
 =
 -\sum_{j=2}^{L-1}\hat{Z}_{j-1}\hat{X}_j\hat{Z}_{j+1}
 -h\sum_{j=1}^{L}\hat{X}_j ,
 \label{eq:s-zxzham}
\end{equation}
with the ZXZ interaction as the energy unit, $J=1$, so the plotted $h/J$ equals the field $h$ used here and in the main text. Choose the ground state with both sublattice parities fixed to $+1$. On every even site $z$, apply $\rho\mapsto(1-p_{\rm e})\rho+p_{\rm e}\hat{X}_z\rho\hat{X}_z$ at $p_{\rm e}=1/2$. The resulting complete $\hat{X}$ dephasing gives
\begin{equation}
 \rho_h
 =
 \sum_s
 P_h(s)\,
 |s\rangle\langle s|_{\rm e}
 \otimes
 |\phi_h(s)\rangle\langle\phi_h(s)|_{\rm o},
 \label{eq:s-zxz-block}
\end{equation}
where $s=(s_1,\ldots,s_{L/2})$ labels the even-sublattice $\hat{X}$ configuration with $\prod_ns_n=+1$, and each conditional odd-sublattice state is pure. Both sublattices remain in the mixed state of Eq.~\eqref{eq:s-zxz-block}; no sites are traced out.

Odd-subchain coordinates $n<m$ correspond to the original-chain sites $i=2n-1$ and $j=2m-1$, separated by $r=|i-j|=2|n-m|$. The indices of $C_{nm}$ in the main text therefore label the odd subchain. Define $\hat{G}=\hat{Z}_i\hat{Z}_j$ and
\begin{equation}
 c_s
 =
 \langle\phi_h(s)|\hat{Z}_i\hat{Z}_j|\phi_h(s)\rangle.
\end{equation}
The blocks in Eq.~\eqref{eq:s-zxz-block} are orthogonal, and the pair rotation preserves their classical weights. Hence
\begin{align}
 Q[\rho_h;\hat{G}]
 &=4\sum_sP_h(s)\left(1-c_s^2\right),\nonumber\\
 D_s(i,j)
 &=\sum_sP_h(s)c_s^2,\nonumber\\
 F_\mathcal O(i,j)
 &=\sum_sP_h(s)|c_s|,\nonumber\\
 R_1(i,j)
 &=\sum_sP_h(s)c_s^2 .
 \label{eq:s-zxz-blockmeans}
\end{align}

To relate these branch expectations to one Ising correlator, define the conditional unitary transformation
\begin{equation}
 q_n(s)=\prod_{k<n}s_k,
 \qquad
 \hat{V}_s=\prod_{n:q_n(s)=-1}\hat{X}_{2n-1}.
\end{equation}
This transformation maps every conditional odd-sublattice problem to the same open transverse-field Ising chain. With its correlator defined by
\begin{equation}
 C_{nm}(h)
 =
 \langle\psi_{\rm I,o}(h)|
 \hat{Z}_{2n-1}\hat{Z}_{2m-1}
 |\psi_{\rm I,o}(h)\rangle ,
\end{equation}
the branch expectation is
\begin{equation}
 c_s
 =
 q_n(s)q_m(s)C_{nm}(h)
 =
 \left(\prod_{k=n}^{m-1}s_k\right)C_{nm}(h).
 \label{eq:s-zxz-sign}
\end{equation}
All branches have the same correlation magnitude. Equations~\eqref{eq:s-zxz-blockmeans} and \eqref{eq:s-zxz-sign} then give the exact finite-size identities
\begin{equation}
 D_s=C_{nm}^2
 =F_\mathcal O^2
 =R_1
 =R_2,
 \qquad
 Q=4(1-C_{nm}^2).
 \label{eq:s-zxzexact}
\end{equation}
Here $R_2=\Tr(\rho_h\hat{G}\rho_h\hat{G})/\Tr(\rho_h^2)$ is the purity-normalized R\'enyi-2 correlator. Its equality with the other quantities follows from this block structure and is not a general hierarchy relation.

Ordinary charged order is determined by the full mixed-state correlator, which contains an additional even-sublattice disorder string,
\begin{equation}
 \langle \hat{Z}_i\hat{Z}_j\rangle_{\rho_h}
 =
 C_{nm}(h)
 \left\langle
 \prod_{k=n}^{m-1}\hat{X}_{2k}
 \right\rangle_{\psi_{\rm I,e}(h)} .
 \label{eq:s-zxz-ordinary}
\end{equation}
For $0<h<1$, the odd-chain correlator approaches
\begin{equation}
 C_\infty(h)=(1-h^2)^{1/4},
\end{equation}
while the even-chain disorder string decays with separation. The ordinary charged correlator therefore vanishes, whereas
\begin{equation}
 D_s^\infty(h)=C_\infty(h)^2=\sqrt{1-h^2}>0,
 \qquad 0<h<1,
 \label{eq:s-zxz-thermolimit}
\end{equation}
which realizes SW-SSB. At the Ising critical point $h=1$, $C_{nm}\sim r^{-1/4}$ gives $D_s\sim r^{-1/2}$.

Finite-size calculations use the free-fermion covariance matrix of the length-$L/2$ Ising chain, with original-chain sizes $L=32,64,128,256$ and separation $r=L/2$. Small-system exact diagonalization checks the conditional mapping, parity assignment, and endpoint indexing. Correlations below the numerical resolution threshold are recorded as unresolved, not as exact zeros.

\begin{figure}[t]
 \centering
 \includegraphics[width=0.88\textwidth]{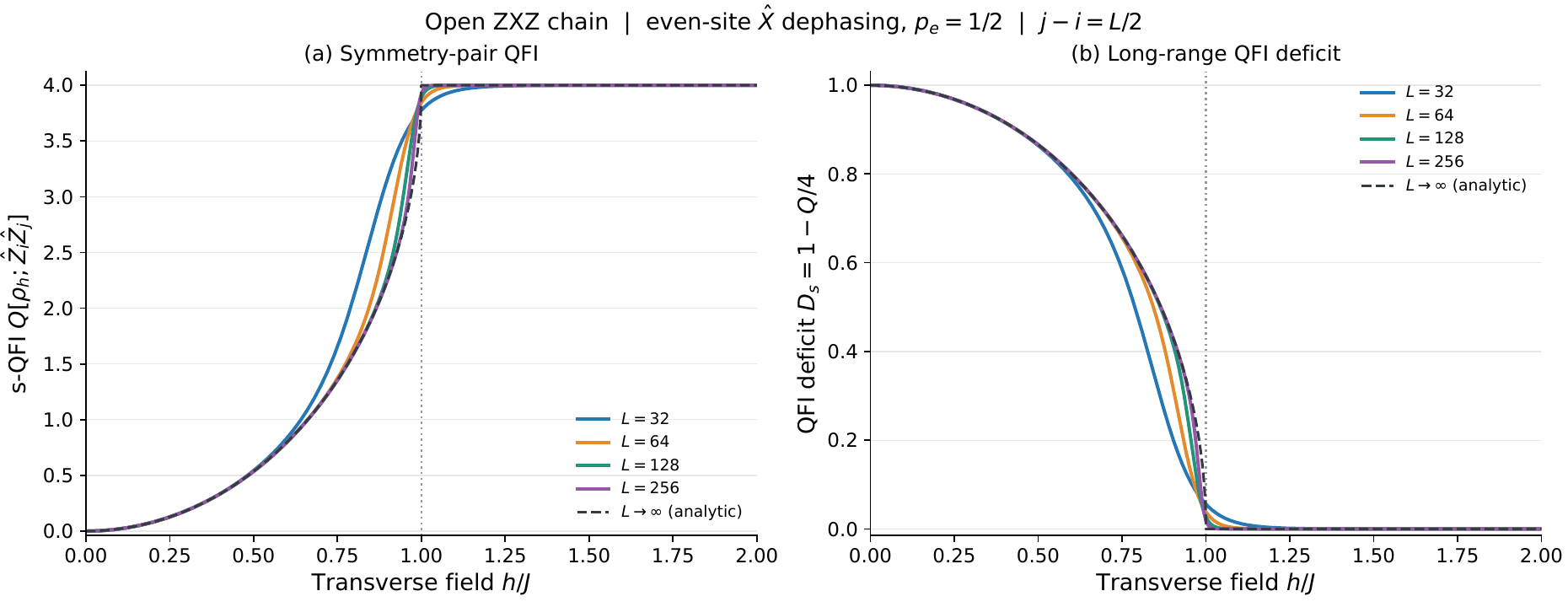}
 \caption{\label{fig:s-zxz}Pair QFI and deficit of the open ZXZ chain after complete even-sublattice dephasing. Solid curves show finite-size calculations; gray dashed curves give the thermodynamic analytic results. The vertical line at $h=1$ marks the exact Ising critical point and is not a fitted crossing.}
\end{figure}

\section{S6. Two-dimensional edge dephasing and RBIM mapping}

Consider an open $L\times L$ square lattice initialized in
\begin{equation}
 \rho_0=|+\rangle\langle+|^{\otimes N},
 \qquad N=L^2,
\end{equation}
with an independent nearest-neighbor edge channel
\begin{equation}
 \mathcal E_{\ell,p}(\rho)
 =
 (1-p)\rho+p\,\hat{Z}_i\hat{Z}_j\rho \hat{Z}_i\hat{Z}_j,
 \qquad \ell=\langle ij\rangle .
 \label{eq:s-edgechannel}
\end{equation}
The resulting state is diagonal in the $\hat{X}$-configuration (syndrome) basis,
\begin{equation}
 \rho_p
 =
 \sum_XP_X|X\rangle\langle X|,
 \qquad
 P_X
 =
 \sum_{s:\,\partial s=X}
 p^{k(s)}(1-p)^{N_e-k(s)},
 \label{eq:s-rhop}
\end{equation}
where $N_e$ is the number of edges, $s$ is an edge-error configuration with $k(s)$ errors, and $\partial s=X$ is its syndrome. Thus $P_X$ is the total probability of syndrome $X$ in the normalized distribution. Each edge error flips two endpoints, enforcing $\prod_iX_i=1$.

The pair $\hat{G}=\hat{Z}_i\hat{Z}_j$ maps $|X\rangle$ to $|X^{ij}\rangle$ by flipping the endpoint charges. As in the main text, omit configurations with $P_X+P_{X^{ij}}=0$ from the weighted sums. The spectral formula gives
\begin{equation}
 D_s(i,j)
 =
 \sum_X
 \frac{2P_XP_{X^{ij}}}{P_X+P_{X^{ij}}},
 \qquad
 F_\mathcal O(i,j)
 =
 R_1(i,j)
 =
 \sum_X\sqrt{P_XP_{X^{ij}}}.
 \label{eq:s-2dmeans}
\end{equation}
The common diagonal basis gives $F_\mathcal O=R_1$, while $D_s$ is generally distinct from both. This contrasts with the exact saturation in the ZXZ chain.

For numerical evaluation, group configurations into unordered pairs $\mathcal P=\{X,X^{ij}\}$ and define
\begin{equation}
 q_{\mathcal P}=P_X+P_{X^{ij}},
 \qquad
 \Delta_{\mathcal P}
 =
 \log\frac{P_X}{P_{X^{ij}}}.
\end{equation}
Omit pairs with $q_{\mathcal P}=0$. If one probability vanishes, interpret $\Delta_{\mathcal P}=\pm\infty$ by continuity. The pair orientation is immaterial because the functions below are even. With $\sum_{\mathcal P}q_{\mathcal P}=1$,
\begin{equation}
 \begin{aligned}
 D_s&=\sum_{\mathcal P}q_{\mathcal P}
 \sech^2\!\left(\frac{\Delta_{\mathcal P}}2\right),\\
 F_\mathcal O=R_1&=\sum_{\mathcal P}q_{\mathcal P}
 \sech\!\left(\frac{\Delta_{\mathcal P}}2\right),\\
 \frac{Q}4&=\sum_{\mathcal P}q_{\mathcal P}
 \tanh^2\!\left(\frac{\Delta_{\mathcal P}}2\right).
 \end{aligned}
 \label{eq:s-bounded}
\end{equation}
Each estimand is bounded. The paired measure and a directed average over $X$ weighted by $P_X$ use different counting conventions, which must be kept consistent in the estimator.

For the random-bond Ising mapping, choose a reference error chain $s^0$ with boundary $X$. Any chain with the same boundary differs by a closed loop. Introduce dual Ising spins $\tau_a=\pm1$, bond signs $\mu_{ab}=(-1)^{s^0_\ell}$, and the random-bond Ising partition function $Z_\mu(\beta_p)$ at coupling $\beta_p$. Then
\begin{equation}
 P_X
 =
 [p(1-p)]^{N_e/2}\,
 Z_\mu(\beta_p),
 \qquad
 \beta_p=\frac12\log\frac{1-p}{p},
 \label{eq:s-rbim}
\end{equation}
with the exterior dual face fixed to remove the global spin-flip redundancy. Flipping endpoints $i,j$ reverses the bond signs along a connecting path $\gamma_{ij}$. Let $\Delta\mathcal F_\mu(i,j)$ be the dimensionless free-energy difference between the endpoint-defect and reference bond configurations. Their probability ratio is
\begin{equation}
 \frac{P_{X^{ij}}}{P_X}
 =
 \frac{Z_{\mu^{\gamma_{ij}}}(\beta_p)}
 {Z_\mu(\beta_p)}
 =
 e^{-\Delta\mathcal F_\mu(i,j)} .
 \label{eq:s-defectratio}
\end{equation}
Paths with the same endpoints differ by a closed gauge deformation and yield the same ratio.

Production calculations use open lattices $L=12,24,36,40$ at $r=L/2$, with $205$ values of $p$ after the low-$p$ supplementation. Each interior point uses $13312$ independent disorder samples and $2000$ bootstrap resamples for pointwise statistical intervals. Main-text Figure~2(c) shows $L=12,24,40$ for clarity. The endpoints $p=0$ and $p=1/2$ are inserted analytically. The sparse Kac--Ward log-determinant backend is cross-checked against small-system exact enumeration or transfer-matrix calculations. The data illustrate the finite-size separation of $F_\mathcal O^2$, $D_s$, and $R_1$; no new thermodynamic critical point is inferred from a curve crossing.

\begin{figure}[t]
 \centering
 \includegraphics[width=0.88\textwidth]{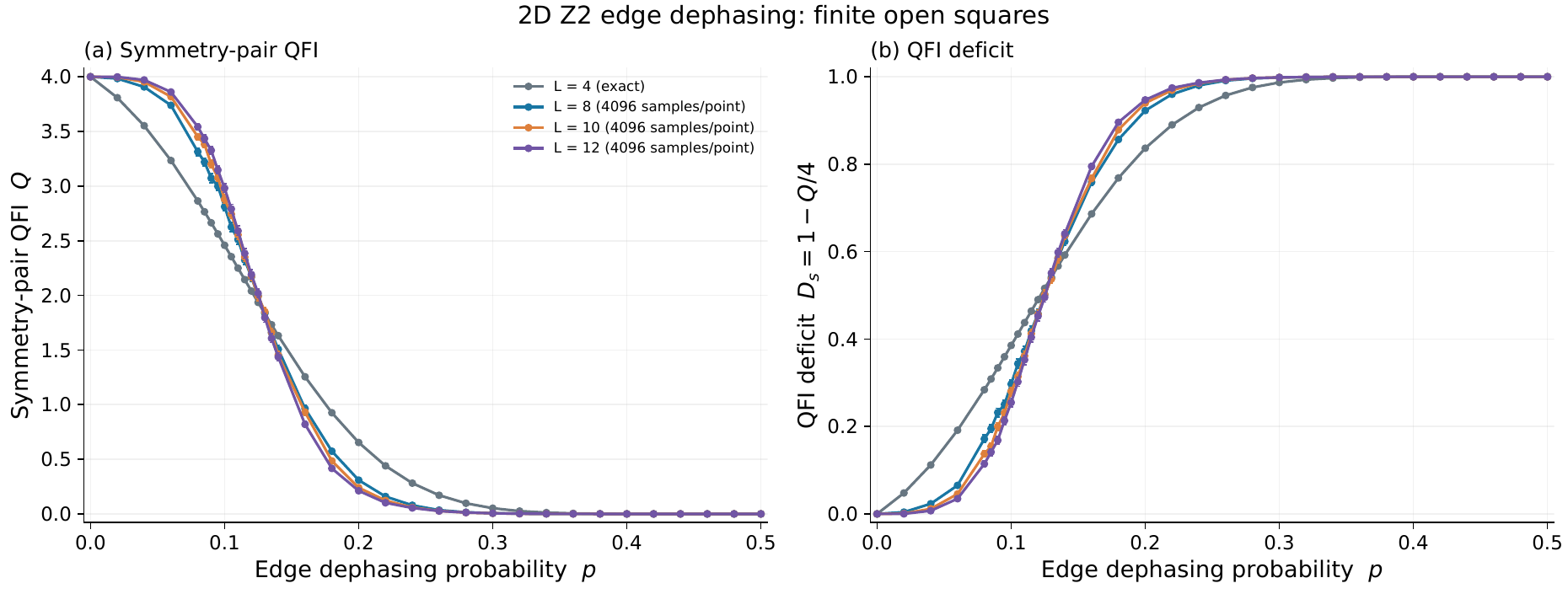}
 \caption{\label{fig:s-overview}Small-system validation of the two-dimensional edge-dephasing calculation. The production analysis uses the larger lattices and sampling protocol described in the text; these small-system curves are not used to locate a thermodynamic transition.}
\end{figure}

\end{document}